\documentclass[preprint]{aastex}

\usepackage{epsfig}
\usepackage[]{natbib}

\newcommand{\eg}{{\rm e.g.}}
\newcommand{\ie}{{\rm i.e.}}
\newcommand{\Mo}{\ensuremath{M_\odot}}
\newcommand{\Morat}{\ensuremath{M/M_\odot}}
\newcommand{\Lo}{\ensuremath{L_\odot}}
\newcommand{\Lorat}{\ensuremath{L/L_\odot}}
\newcommand{\Bo}{\ensuremath{B_\odot}}

\newcommand{\Lstar}{\ensuremath{L_{\star}}}

\newcommand{\Bstar}{\ensuremath{B_{\star}}}

\newcommand{\amin}{\ensuremath{\arcmin}}
\newcommand{\asec}{\ensuremath{\arcsec}}
\newcommand{\ml}{\ensuremath{M/L}}
\newcommand{\mlb}{\ensuremath{M/L_{B}}}
\newcommand{\mlv}{\ensuremath{M/L_{V}}}
\newcommand{\mlo}{\ensuremath{\Mo/\Lo}}
\newcommand{\hmlo}{~\ensuremath{{h}{\mlo}}}
\newcommand{\kmsMpc}{~\ensuremath{{\rm km\ sec}^{-1}\ {\rm Mpc}^{-1}}}

\newcommand{\hkpc}{~\ensuremath{{h}^{-1}{\rm kpc}}}
\newcommand{\hMpc}{~\ensuremath{{h}^{-1}{\rm Mpc}}}

\newcommand{\Hnought}{\ensuremath{{\rm H}_0}}
\newcommand{\znought}{\ensuremath{z_0}}

\newcommand{\omegamnought}{\ensuremath{\Omega_{m0}}}
\newcommand{\omegalnought}{\ensuremath{\Omega_{\lambda0}}}
\newcommand{\scrit}{\ensuremath{\Sigma_{\rm crit}}}
\newcommand{\scritrat}{\ensuremath{\Sigma/\Sigma_{\rm crit}}}
\newcommand{\scritinv}{\ensuremath{1/\Sigma_{\rm crit}}}

\def \figwidth {\linewidth}

\shorttitle{Title}
\shortauthors{Wilson \etal}

\begin{document}

\title{MASS AND LIGHT IN THE UNIVERSE\altaffilmark{1}}

\author{Gillian Wilson\altaffilmark{2,3}, Nick Kaiser\altaffilmark{2} and Gerard A. Luppino\altaffilmark{2}}

\email{gillian@het.brown.edu}

\altaffiltext{1}{Based on observations with the Canada-France-Hawaii Telescope which is operated by the National Research Council of
Canada, le Centre National de la Recherche Scientifique de France, and the University of Hawaii.}
\altaffiltext{2}{Institute for Astronomy, University of Hawaii, 2680 Woodlawn Drive, Honolulu, HI 96822}
\altaffiltext{3}{Physics Department, Brown University, 182 Hope Street, Providence, RI 02912}

\begin{abstract}
We present a weak lensing and photometric study of 
six $0^{\circ}.5 \times 0^{\circ}.5$ degree
fields observed at the CFHT using the UH8K CCD mosaic camera.
The fields were observed for a total of 2 hours each in $I$ and $V$,
resulting in catalogs containing $\sim 20$,$000$ galaxies per passband per field.
We use $V-I$ color and $I$ magnitude to select bright early type galaxies 
at redshifts $0.1 < z < 0.9$. 
We measure the gravitational shear from faint galaxies in the range $21 < m_{I} < 25$
from a composite catalog and find a strong correlation with that predicted from the early types
if they trace the mass with $\mlb \simeq 300\pm75 \hmlo$ for a flat ($\Omega_{{\rm m}0} = 0.3, \Omega_{\lambda 0} = 0.7$) lambda cosmology and
$\mlb \simeq 400\pm100 \hmlo$ for Einstein-de Sitter.
We make two-dimensional reconstructions of 
the mass surface density. Cross-correlation of the measured 
mass surface density with that predicted from the
early type galaxy distribution shows a strong peak at zero lag (significant at the $5.2\sigma$
level). We azimuthally average the cross- and auto-correlation functions. 
 We conclude that the
profiles are consistent with early type galaxies tracing  mass 
on scales of
$\geq45\asec$ ($\geq200 \hkpc$ at $z=0.5$). 
We sub-divide our bright early type galaxies by redshift and obtain similar conclusions.
These $\mlb$ ratios imply $\omegamnought \simeq 0.10\pm0.02$ ($\omegamnought \simeq 0.13\pm0.03$ for Einstein-de Sitter) of closure density.

\end{abstract}


\section{INTRODUCTION}
\label{sec:intro}

It is well known that a large quantity of dark matter exists in the Universe.
Evidence for dark matter around luminous galaxies
comes from stellar velocity dispersions  and
rotation curves in the outer parts of spiral 
galaxies \markcite{fg-77,bos-81, trim-87}({Faber} \& {Gallagher} 1979; {Bosma} 1981; {Trimble} 1987); and large
velocity dispersions \markcite{fg-77,trim-87}({Faber} \& {Gallagher} 1979; {Trimble} 1987)
and extended X-ray halos of hot gas \markcite{mush-94, trinch-94, kf-95, trinch-97}({Mushotzky} {et~al.} 1994; {Trinchieri} {et~al.} 1994; {Kim} \& {Fabbiano} 1995; {Trinchieri}, {Fabbiano}, \&  {Kim} 1997) in elliptical galaxies.
On larger scales,  evidence for dark matter in clusters comes from
gravitational lensing \markcite{mel-99}({Mellier} 1999, and references therein), virial analyses 
\markcite{carl-96}({Carlberg} {et~al.} 1996), or 
X-ray halos of hot gas \markcite{wf-95}({White} \& {Fabian} 1995).
Evidence for dark matter in the field comes from 
relative motions of galaxies in the Local Group \markcite{tur-76, san-86, jmb-98}({Turner} 1976; {Sandage} 1986; {Jing}, {Mo}, \& {Boerner} 1998),
or
relative motions of
faint satellites \markcite{bt-81,zar-97}({Bahcall} \& {Tremaine} 1981; {Zaritsky} {et~al.} 1997),
or 
pairs of galaxies analyzed statistically \markcite{tur-76, bp-87, dmw-97, jmb-98}({Turner} 1976; {Brown} \& {Peebles} 1987; {Davis}, {Miller}, \& {White} 1997; {Jing} {et~al.} 1998).
On still larger scales of $0.25-3 \hMpc$, evidence for dark matter comes from the
cosmic virial theorem analysis \markcite{dp-83}({Davis} \& {Peebles} 1983) and least action method \markcite{shaya-95}({Shaya}, {Peebles}, \& {Tully} 1995), and
on $10-30 \hMpc$ scales, from bulk flows and redshift-space anisotropies \markcite{sw-95}({Strauss} \& {Willick} 1995, and references therein).

The relative contribution of the dark matter component is usually specified in terms of the 
mass-to-light ratio, $\ml$, the ratio of the total mass relative to the total light within
a given scale. It is generally acknowledged that
the $\ml$ ratio increases from the bright luminous regions
of galaxies to their faint halos, with possible further increase on larger scale to systems such
as groups and rich clusters of galaxies. The first measurement of 
the $\ml$
ratio in the Coma cluster \markcite{zw-33}(Zwicky 1933) obtained  $\ml\sim300 \hmlo$. Subsequent measurements of a series of clusters have confirmed
his original numbers (\markcite{carl-97}{Carlberg} {et~al.} 1997 find a virial $\ml =213\pm59$ for galaxy clusters assuming an $\omegamnought = 0.2, \omegalnought = 0.0$ cosmology - see also \markcite{carl-96}{Carlberg} {et~al.} 1996). If the Coma $\ml$ ratio is universal, then the density parameter
of the Universe would appear to be $\omegamnought \simeq 0.2$. 
If one wished to reconcile
cluster $\ml$ ratios with the philosophically appealing value of 
$\omegamnought = 1$  one was forced to argue that
the efficiency of galaxy formation must therefore be biased (enhanced) in dense environments \markcite{kais-84, bbks-86}({Kaiser} 1984; {Bardeen} {et~al.} 1986).
As one measured $\ml$ on larger and larger scales one might expect 
the $\ml$  ratio to increase
until one approached the true global value of $\omegamnought = 1$. Motivated 
by such reasoning,
much effort has been expended, both in simulating bias
on galaxy, cluster or large-scale structure scales \markcite{defw-85}({Davis} {et~al.} 1985) and also in 
attempting to measure its presence from large-scale galaxy bulk flows
\markcite{sigad-98, ws-98, branch-00}({Sigad} {et~al.} 1998; {Willick} \& {Strauss} 1998; {Branchini} {et~al.} 2000).
For a time, the idea that $\ml$ ratios increased as a function of increasing scale seemed very plausible.
A very clear summary of $\ml$ ratio with scale from a variety of methods is given in \markcite{bld-95}{Bahcall}, {Lubin}, \& {Dorman} (1995, hereafter BLD). In
that paper, however, it is argued, that while $\ml$ increases with scale
to $\simeq 200 \hkpc$, there is little evidence that $\ml$ ratios increase on scales beyond that. BLD
argued that the total mass of large-scale systems such as groups, rich clusters and superclusters could
be accounted for by the total mass of their member galaxies including their large halos and intracluster gas.
They argued for a $\mlb\simeq100\hmlo$ for late type galaxies and a $\mlb\simeq400\hmlo$ for early type galaxies
and concluded that these values implied  $\omegamnought \sim 0.2-0.3$. 

Strong evidence from \markcite{kwlkgmd-01}Kaiser {et~al.} (2001c, hereafter KWLKGMD) also suggested a similar picture. In that paper, ``A Photometric and 
Weak Lensing Analysis of the z = 0.42 Supercluster ms0302+17'', it was shown 
on scales $\geq 200\hkpc$
that \emph{early} type galaxy light traces mass with $\mlb \simeq 250\pm50 \hmlo$. This was the first time that
mass had been measured out to such a large radial distance ($\simeq3\hMpc$) from a cluster center  using such a ``direct'' technique 
as gravitational lensing.
This ``light-traces-mass'' relationship was somewhat surprising and intriguing. It raised the question of 
whether the relationship (and indeed $\mlb$ value) was exclusively applicable to early type galaxies in rich
cluster environments or was applicable to early type galaxies in all environments in the Universe.

In this paper we investigate the relationship between mass and luminosity
on scales of up to $30\amin$ using data collected at the CFHT
with the UH8K camera.  Our analysis differs from previous lensing studies in that
here we focus on ``blank fields'' \ie\ the fields chosen for study were intended to
be representative views of the universe which
do not contain any unusually large masses such as rich clusters. 
We investigate the same hypothesis as proposed in KWLKGMD - 
namely that early type galaxy light traces mass with a constant ratio of proportionality.

The outline of the paper is as follows.  In \S\ref{sec:data} we describe the
data and the selection of lens and background galaxies. We also present surface mass density reconstructions from shear estimates.
In \S\ref{sec:ml} we 
compare these to predictions inferred from the luminosity of early type galaxies at various redshifts.
In \S\ref{sec:disc} we discuss our results.  We 
calculate the mean mass-to-light ratio of an early type galaxy and the
contribution of early types to the closure density. We investigate the
dependence of these values on cosmology.
We then compare our values of $\mlb$ and $\omegamnought$ with  
other studies. We also consider possible sources of uncertainty. 
In \S\ref{sec:conc}  we briefly summarize our conclusions.
We assume a 
flat lambda ($\Omega_{{\rm m}0} = 0.3, \Omega_{\lambda 0} = 0.7$) 
cosmology with $\Hnought = 100 $ $h\kmsMpc$ throughout unless
explicitly stated otherwise. 

\section{THE DATA AND GALAXY SAMPLES}
\label{sec:data}

\subsection{Data Acquisition and Reduction}

The data were taken at the 3.6m CFHT telescope using the $8192 \times 8192$
pixel UH8K camera at prime focus. The field of view of this camera is 
$\simeq 30 \amin$ with pixelsize $\simeq 0\asec.207$. The data (six pointings) used in the analysis were acquired as part
of an ongoing 
project whose principle aim is to investigate the cosmic shear pattern caused
by gravitational lensing from the large-scale structure  of the Universe.
Table~\ref{tab:fields} gives an overview of the data, describing the field 
name, center and seeing for each pointing. 
This is the third in a series of papers describing results from the 
large-scale structure project.
\markcite{kwl-01}Kaiser, Wilson, \&  Luppino (2001a, Paper I, hereafter KWL) presented estimates of cosmic shear variance 
on $2' - 30'$ scales, and \markcite{wklc-01}Wilson {et~al.} (2001b, Paper II, hereafter WKLC) investigated galaxy halos 
at radii of $20'' - 60''$ ($50 -200\hkpc$).
Here we focus on mass and light on galaxy group and cluster scales.
A forthcoming paper will address galaxy
clustering. Further details of the data reduction pipeline may be 
found in \markcite{kwld-01}Kaiser {et~al.} (2001b),
and, as already mentioned in the Introduction, an application to the ms0302 supercluster in KWLKGMD.
In brief, the data were dark subtracted, flat-fielded, registered, median averaged
and corrected for galactic extinction.
A full description
of our catalogs will be presented in a later paper \markcite{wkcats-01}(Wilson \& Kaiser 2001).
 
\begin{deluxetable}{ccccrrcc}
\tablewidth{0pt}
\tablecaption{Field Centers and  Seeing 
\label{tab:fields}
}
\tablehead{
\colhead{Field} &
\colhead{Pointing} &
\colhead{RA (J2000)} &
\colhead{DEC (J2000)} &	
\colhead{l } &	
\colhead{b } &	
\colhead{FWHM(I)} &	
\colhead{FWHM(V)} 	
}
\startdata
Lockman	& 1	& 10:52:43.0	& 57:28:48.0 	& $149.28$	& $53.15$	& $0''.83$	& $0''.85$	\\
	& 2	& 10:56:43.0	& 58:28:48.0 	& $147.47$	& $52.83$ 	& $0''.84$	& $0''.86$ 	\\
Groth	& 1	& 14:16:46.0	& 52:30:12.0	& $96.60$	& $60.04$	& $0''.80$	& $0''.93$	\\
	& 3	& 14:09:00.0	& 51:30:00.0 	& $97.19$	& $61.57$	& $0''.70$	& $0''.85$	\\
1650	& 1	& 16:51:49.0	& 34:55:02.0 	& $57.37$	& $38.67$	& $0''.82$	& $0''.85$	\\
	& 3	& 16:56:00.0	& 35:45:00.0 	& $58.58$	& $37.95$	& $0''.85$	& $0''.72$	\\
\enddata
\end{deluxetable}







\subsection{Lens Galaxy Sample}
\label{ssec:fg}

Our analysis differs from other approaches in that we use $V -I$ color
to select a sample of bright early type lens galaxies with reasonably
well determined redshifts.  As we will show later, these trace
the mass and thus by focusing on the distribution
of early type galaxies one can accurately forecast the distribution
of mass in the Universe.

As shown in section 2.2 of  WKLC, with
fluxes in 2 passbands one can reliably select bright early
type galaxies and assign them approximate redshifts.  This is because
early type galaxies are the reddest galaxies at a given redshift.  Thus, if
we select galaxies of some color $c$ we will see a superposition
of early types at redshift $z_E$ such that $c = c_E(z_E)$
and later types at their appropriate, but considerably higher, redshift.  
An $L \sim L_\star$ early
type galaxy 
will appear much brighter than an $L \sim L_\star$ late type galaxy 
by about 3 magnitudes,
so with
a judicious cut in red flux it is possible to isolate a 
bright early type
galaxy sample. At lens redshift higher than $z \simeq 0.4$, all galaxies with colors $V-I>2.4$ are
early types and it is unnecessary to apply any magnitude cut to exclude late types. We do, however,
exclude galaxies with $m_{I}$ fainter than $23.0$.  
Figure 2 of WKLC shows $V-I$ color versus redshift for four galaxy types and
figure 1 from the same paper shows counts predicted for all galaxy types
and also the specific magnitude cut we employ at each redshift to ensure that only early types remain in our sample.
Table~\ref{tab:lens} shows the number of early type lens galaxies per redshift slice ($dz = 0.1$)
brighter than the magnitude cut (summed over all six pointings).

\begin{deluxetable}{crc}
\tablewidth{0pt}
\tablecaption{Lens Galaxy Data.
\label{tab:lens}
}
\tablehead{
\colhead{Lens Redshift} &
\colhead{Number Lens} &
\colhead{$\Sigma_{\rm crit}^{-1} (\times10^{-16} {\hMpc}^{2} \Mo^{-1})$\tablenotemark{a}}
}
\startdata
$0.1\pm0.05$  & $92$   & $1.36$
\\
$0.2\pm0.05$  & $222$  & $2.04$
\\
$0.3\pm0.05$  & $366$  & $2.28$
\\
$0.4\pm0.05$  & $960$  & $2.26$
\\
$0.5\pm0.05$  & $1611$ & $2.10$
\\
$0.6\pm0.05$  & $663$  & $1.86$
\\
$0.7\pm0.05$  & $699$  & $1.61$
\\
$0.8\pm0.05$  & $594$  & $1.36$
\\
$0.9\pm0.05$  & $233$  & $1.13$
\\
\enddata
\tablenotetext{a}{Cosmology dependent ($\Omega_{{\rm m}0} = 0.3, \Omega_{\lambda 0} = 0.7$ assumed here).}
\end{deluxetable}

\subsection{Background (Source) Galaxy Sample}
\label{ssec:bg}

The background sample was selected to lie in a range of
significance $4 < \nu < 150$ (equivalent to limiting magnitudes of
$m_{I} \simeq 25$ and $m_{I} \simeq 21$ for a 
point source). Shear estimates for each galaxy,  
$\hat{\gamma}_{\alpha}$ (for $\alpha = 1,2 $), were determined 
using the 
method described in \markcite{kais-00}{Kaiser} (2000) and KWL. Weighted second moments 
were calculated from 

\begin{equation}
\label{eq:qalpha}
{q}_{\alpha} = M_{\alpha lm} \int d^2 r \; S(r) r_{l} r_{m} f( \mathbf{r})
\end{equation} 
where
$f$ is flux;
$r$ is projected 
angular separation from the galaxy center;  
$S$ is a Gaussian smoothing function to prevent the integral diverging at large radii;
and the two constant
matrices $M_1$ and $M_2$ are 
\begin{equation}
\label{eq:matrices}
M_{1lm} \equiv \left[ 
\begin{array}{cc}
1 & 0\\
0 & -1\\
\end{array}\right],
\;
M_{2lm} \equiv \left[ 
\begin{array}{cc}
0 & 1\\
1 & 0\\
\end{array}\right].
\end{equation} 
Weighted second moment shapes and magnitudes of objects were measured
using varying aperture photometry. The final number of galaxies per 
pointing and passband
is shown in Table~\ref{tab:cats}.

\begin{deluxetable}{ccrrr}
\tablewidth{0pt}
\tablecaption{Galaxy Catalogs.
\label{tab:cats}
}
\tablehead{
\colhead{Field} &
\colhead{Pointing} &
\colhead{I} &
\colhead{V} &	
\colhead{IV} 
}
\startdata
Lockman	& 1	& 20820 & 20358 & 25963 \\
	& 2	& 20428 & 17782 & 23835 \\
Groth	& 1	& 27906 & 16391 & 29437 \\
	& 3	& 19300 & 15876 & 22989 \\
1650	& 1	& 21785 & 15403 & 24494 \\
	& 3	& 18391 & 16518 & 22894 \\
\enddata
\end{deluxetable}

A `best' combined $IV$ catalog was also created. This is a catalog containing 
galaxies which have been detected in \emph{both} $I$ and $V$ images
above a  threshold significance (of $4\nu$). This is to ensure 
that any given ``detection'' is truly a real object.
Shape information, \ie\ shear estimates, are retained from the higher 
significance passband detection and discarded from the alternate passband. The galaxies tend
to be detected at higher significance in the $I$-band images and we find that
the majority ($\simeq 80\%$) of galaxies
in the combined $IV$ catalog originate from the $I$ catalog. The final number of objects 
in each $IV$ catalog is shown in Table~\ref{tab:cats}. 
(As discussed in KWL
there are some low-level systematics still present in the catalogs. However, these
are likely to have very little effect on the results presented in this paper for two 
reasons. Firstly, in this paper we analyze the light-mass cross-correlation rather than
the mass auto-correlation investigated in KWL. Thus any systematic component to the
shear will not correlate
with the light, and is likely to average out. Secondly, we utilize mainly the 
$IV$ catalog and this contains mostly galaxies originating from the
$I$-band which was shown in KWL to be less affected by systematics than the 
$V$-band).

From galaxy shear estimates, $\hat{\gamma}_{\alpha}$, we constructed two-dimensional mass surface density 
reconstructions in terms of the dimensionless quantity $\kappa$ (where $\kappa = \scritrat$,
the physical mass per unit area in units of the critical surface density). 
For any given lens and source galaxy redshift the critical
surface density, $\scrit$, is
the mass surface density required to refocus light. In the case
of a distribution of background galaxy redshifts, 
$\scrit$ becomes an average or effective mass surface density,
and is given by

\begin{equation}
\label{eq:scritinv}
\scritinv  = \frac {4 \pi G}{c^2} \frac {a_{0} \omega_{l}}{1 + z_{l}} \langle \beta(z_{l}) \rangle,
\end{equation}
where $\omega$ is comoving distance measured in units of the current curvature scale
$a_0 = c / (H_0 \sqrt{1 - \omegamnought - \omegalnought})$ and
the dimensionless quantity $\langle \beta (z_{l}) \rangle$ is
defined as 
\begin{equation}
\label{eq:betaeff}
\langle \beta (z_{l}) \rangle \equiv \frac{\int^{\infty}_{0} dz_s \; n_s(z_s) 
\langle W_s(z_s) \rangle \beta(z_l, z_s)}
{\int^{\infty}_{0} dz_s \; n_s(z_s) \langle W_s(z_s) \rangle}
\end{equation}
where $n_s(z)$ is the redshift distribution of the source galaxies,
$\langle W_s(z_s) \rangle$ is the mean weight for source galaxies at
redshift $z_s$,
and where, finally,
\begin{equation}
\label{eq:betalambda}
\beta(z_l, z_s) \equiv {\rm max}(0,\sinh{(\omega_{s}-\omega_{l})}/\sinh{(\omega_{s})}) 
\end{equation}
Physically, $\beta(z_l, z_s)$ is the ratio of the distortion induced by
a lens at redshift $z_l$ in
an object at
finite distance $\omega(z_s)$ relative to that for a fictitious source at infinite
distance.

For the special case of a spatially flat cosmology, $\omega \rightarrow 0$ and
$a_0 \rightarrow \infty$, but such that their product remains finite.  In that case
$\sinh \omega \rightarrow \omega$, and
$\langle \beta \rangle \equiv  \langle {\rm max}(0,1 - \omega_{l}/\omega_{s})\rangle$.
For the limiting case of $\Omega_m = 1$, $\Omega_\lambda = 0$,
$\omega(z) = 2(1 - 1/ \sqrt{1 + z})$ and, 
in the other extreme, for  $\Omega_m \rightarrow 0$, $\Omega_\lambda \rightarrow 1$,
$\omega(z) = z$.

Figures~\ref{fig:k_lock} to~\ref{fig:k_16503} show two-dimensional 
reconstructions of $\kappa$
using galaxy shear estimates from the catalogs described in Table~\ref{tab:cats}. The
\markcite{ks-93}Kaiser \& Squires (1993) reconstruction algorithm was used. This is a stable and fast reconstruction method
which has very simply defined noise properties; essentially Gaussian white noise. As with
all reconstruction methods, there is
a tendency for noise to increase near the data boundaries. 
The upper panels show the reconstructions from catalogs made from the $I$ and $V$-band
observations separately. The lower left panel shows the
reconstruction from the composite $IV$ catalog (our preferred catalog), and the lower right panel shows the
reconstruction using a randomized catalog (containing the same galaxy positions as the original $IV$ catalog but with randomly shuffled shear values), indicating the expected
noise fluctuations due to intrinsic random galaxy shapes. The mass reconstruction from the $IV$ catalog is very similar to that from the
$I$ catalog as might be expected if the majority of objects originated from that catalog. 
The reconstructions have been smoothed
with a  $45\asec$ Gaussian filter.
The wedge shows the 
calibration of the grayscale and the contour separation is $0.04\times\scritrat$.

The first thing that one notices about the
mass surface density reconstructions in Figures~\ref{fig:k_lock} to~\ref{fig:k_16503} 
is that mass structures are much more difficult to discern than for 
reconstructions of clusters. At first glance, it is often 
somewhat unclear as to  whether 
any given peak (or trough) in the mass reconstruction is real, or is simply 
a spurious noise feature due to  
discrete sampling of the background wallpaper of galaxies and their intrinsic ellipticities.  
It would be very difficult to 
quantify mass distributions directly from these maps. 

The eye can be
deceiving, however. The $I$ and $V$ reconstructions are actually extremely similar 
with regards to the positioning of the major mass
distributions. 
Figure~\ref{fig:inv} shows the 
cross-correlation of the mass reconstruction from the $V$ catalog with 
the mass reconstruction 
from the $I$ 
catalog (left panel) and with the mass reconstruction from the $I$ catalog with randomized shear
values (right panel)
for Lockman (upper) to 1650 (lower) pointings. In each case, there is a prominent peak at zero lag.
The peak is significant at the $\sim 6-8\sigma$ level, depending on the pointing.
That the $I$- and $V$-band signal is similar does not prove conclusively that 
galaxies in the $I$ and $V$ 
catalogs have been lensed by the same foreground mass structures (after all, the
$I$ and $V$ catalogs contain many of the same objects and might simply be subject to the
\emph{same} systematics) but it is certainly reassuring.

Figure~\ref{fig:hist} illustrates more quantitatively the difficulty of teasing 
true signal from the noise.
This figure shows two histograms of absolute pixel value from the reconstructions of mass surface density $\kappa$ using the shear estimates from all six $IV$ catalogs (Figures~\ref{fig:k_lock} to \ref{fig:k_16503}).
The solid histogram describes originally positive pixels and the dashed histogram describes
originally negative pixels. The inset shows the contrast for extreme values.
Clearly the distribution is very symmetrical. Notably, a highly positive tail is absent,
indicating the absence of very overdense structures \eg\ rich clusters.

In view of the difficulty of measuring mass \emph{directly} from Figures~\ref{fig:k_lock} 
to \ref{fig:k_16503}, and since we have a large area containing many structures, it may be 
possible to better reveal
the signal by cross-correlating light and mass.
In the following section we cross-correlate the luminosity associated 
 with foreground galaxies with the mass inferred from the background galaxy shear estimates.


\section{MASS AND LIGHT}
\label{sec:ml}

\subsection{Mass Surface Density Predictions from Luminosity}
\label{ssec:pred}

In this section we generate predictions of the dimensional mass surface density $\kappa$ 
from $I$-band galaxy luminosity, assuming a constant $\mlb$. The implicit assumption is that 
optical early type galaxy luminosity is an unbiased tracer of the mass.

For a single galaxy with observed magnitude $m_{I}$, and for 
any cosmology, the contribution to an image of $\kappa$ from a galaxy at redshift $z_{l}$ which 
falls in a pixel with solid angle $d\Omega$ is

\begin{equation}
\label{eq:kdomega}
\kappa d\Omega  = \frac{M}{L_{B}} \frac{4 \pi G\Mo a_{0}}{c^2 (10{\rm pc})^2} (1+z_{l})^3 w_{l} \langle \beta(z_{l}) \rangle 10^{0.4(M_{\Bo} - m_{I} +K_{BI}(z_{l}))}
\end{equation}
where $\mlb$ is in solar units \ie\ $\hmlo$. (obtained from 
$\kappa = \case{M/d\rm{A}}{\scrit}$ where A is area, $\Morat = \ml\times\Lorat$, and
$\Lorat = 10^{-0.4(M_{B} - M_{\Bo)}}$ where $M_{\Bo}$ is the absolute magnitude
of the sun in the B-passband and where $K_{BI}(z) = K_{I}(z) - (M_B - M_I)_{0}$ is the combination of the
conventional $K$-correction and the rest frame color).

From equations~\ref{eq:scritinv} and \ref{eq:kdomega}, $\kappa$ is a function of both lens and source galaxy
redshift. The lens redshift is known fairly accurately from the $V-I$ color, but, in order 
to make an accurate prediction for the dimensionless surface density $\kappa$
it is necessary to have an accurate prediction for $\Sigma_{\rm crit}$ and hence to have an accurate 
model for the redshift
distribution of the faint source galaxies. 

The catalogs used
here are not particularly deep, 
and there are nearly complete
redshift samples which probe similar magnitude ranges. In WKLC
the catalogs were compared to the SSA22 field sample from Cowie's (ongoing) galaxy survey 
\markcite{cghskw-94,cshc-96, csb-99, wcbb-01}(Cowie {et~al.} 1994; {Cowie} {et~al.} 1996; Cowie, Songaila, \& Barger 1999; Wilson {et~al.} 2001a).
In both of our $I$- and $V$-band samples the weight is distributed over a range of several
magnitudes, with half of the weight attributed to galaxies 
brighter/fainter than $m_{I} \simeq 23.0$ and $m_{V} \simeq 24.2$.
The very faintest galaxies lie beyond the completion
limit of Cowie's sample, but the redshift distribution in a
band one magnitude wide about the median magnitude is
well determined. To a first approximation, the
effect of variation of mean redshift with magnitude should 
cancel out, so one can adopt the 
central band redshift distribution as appropriate for the full sample.
At this magnitude the samples are approximately 80\% complete,
and it is thought that the galaxies for which a redshift cannot
be obtained lie predominantly around $z = 1.5-2.0$.

The redshift distribution was shown to be well modeled  by
\begin{equation}
\label{eq:pz}
p(z)  = 0.5 z^{2} \exp(-z/\znought) / \znought^{3}
\end{equation}
for which the mean redshift is $\overline{z} = 3 z_0$ and the
median redshift is $z_{\rm median} = 2.67 z_0$.
This is also the analytic form used by \markcite{witt-00}{Wittman} {et~al.} (2000) and
others, and seems to adequately describe the data.
To allow for incompleteness we set the parameters $n_0$, $z_0$
of the model distribution to match the total number of
galaxies in the Cowie sample (with and without secure redshifts) and to match the
mean redshift with the unmeasurable objects assigned
a redshift $z = 1.8$.
Figure 4 of WKLC
shows the redshift distribution 
for galaxies around $m_{I} = 23.0$
along with the incompleteness corrected model, which has
redshift scale parameter $z_0 = 0.39$.  The same calculation for
galaxies selected in a one magnitude wide band around $m_{V}=24.2$
yields a slightly smaller, though very similar, redshift
parameter $z_0 = 0.37$.  
Thus, the $I$ and $V$ catalogs probe to similar depth in redshift.

In Figure~\ref{fig:invscrit} we plot $\scritinv$ as a function of lens
redshift for three cosmologies using equation~\ref{eq:pz} as the source galaxy distribution. The dot-dashed line is flat lambda ($\omegamnought = 0.3, \omegalnought = 0.7$), the solid line is 
Einstein-de Sitter ($\omegamnought = 1.0, \omegalnought = 0.0$), the dashed line is
open baryon ($\omegamnought = 0.05, \omegalnought = 0.0$). The values of $\scritinv$ for the
flat lambda case are shown in Table~\ref{tab:lens} as a function of z for redshift intervals $dz = 0.1$.
We return to the dependence of $\ml$ ratio on cosmology in \S\ref{ssec:cosmo}.

To obtain the total mass surface density along the line of sight out to $z = 1$ 
we use Table~\ref{tab:lens} to calculate $\kappa = \scritrat$ for each redshift slice individually and then
sum the slices together.
One could directly compare $\kappa$ predicted from the light to $\kappa$
from galaxy shear estimates obtained in \S\ref{ssec:bg}. However, to ensure that 
our $\kappa$-from-light prediction
is subject to the same finite-field bias and other unknown systematics
as the $\kappa$-from-shear reconstructions, 
we firstly make a shear field image prediction from the constant $\mlb$ prediction 
and then sample this at the actual positions of our faint galaxies to generate a synthetic
catalog (that which would have been observed with no intrinsic random shape or measurement noise),
and then generate a reconstruction from that synthetic catalog. To match the spatial resolution to 
that of the real reconstructions (a $45\asec$ Gaussian) we generate the predicted shear with smoothing
scale $45\asec/\sqrt{2}$ and create the reconstruction from the synthetic catalog with the same smoothing.
While correctly accounting for the finite field effect on structures within the field, the actual shear
may still feel some effect from structures outside of the field.
 
The upper left panel of Figures~\ref{fig:pred_lock} to~\ref{fig:pred_16503} shows 
dimensionless surface mass density $\kappa$-from-light generated in this manner. 
We assumed a  $\mlb = 300\hmlo$. The mean has been subtracted from each image. 
The wedge shows the calibration of the grayscale
and the contour separation is $0.007\times\scritrat$.


\subsection{Mass-Light Cross-Correlation}
\label{ssec:mlx}

In this section we cross-correlate light with mass. Our aim is firstly to test the hypothesis of a constant
mass-to-light ratio which is independent of scale, and  also later to determine the value of $\mlb$,
the constant of proportionality between mass and light.
That is to say we are comparing the lower left panel of Figures~\ref{fig:k_lock}-\ref{fig:k_16503} with the
upper left panels of Figures~\ref{fig:pred_lock}-\ref{fig:pred_16503}.  We also require an estimate
of uncertainties inherent in the reconstruction of $\kappa$ from the galaxy shear estimates.
To obtain this we create an
ensemble of 32 reconstructions for each pointing using shear values measured from the $IV$ 
catalog, but shuffled randomly.

The left panel of Figure~\ref{fig:ccf1} shows the cross-correlation of light with mass
averaged over all six pointings. In computing this we padded the source images with 
zeros to 
twice the original size.
The right panel of Figure~\ref{fig:ccf1} shows the
cross-correlation of light with an average over  randomized catalog reconstructions.
There is a strong cross-correlation peak at zero lag for the real data, which is not present
for the randomized data.

The mass-to-light ratio, $\mlb$, at zero lag is calculated by minimization
of 

\begin{equation}
\label{eq:chisq}
\chi^{2} = \sum_{i}{{\frac{(y_{i} - \mlb x_{i})}{\sigma_{i}^{2}}}^{2}}
\end{equation}
where the sum is over the six pointings, the light auto-correlation $x_{i} = \Sigma_{\rm pix} \kappa_{l} \kappa_{l}/N_{\rm pix}$, the mass-light cross-correlation $y_{i} = \Sigma_{\rm pix} \kappa_{m} \kappa_{l}/N_{\rm pix}$, and the uncertainty is calculated from the ensemble of 32 randomized catalogs ( $\sigma_{i}^{2} = \sum{(y_{i}^{\rm rand})^{2}}/ N_{\rm rand}$).
Minimizing $\chi^{2}$ with respect to $\mlb$,

\begin{equation}
\label{eq:mlrat}
\mlb = \frac{\sum_{i}{x_{i}y_{i}/\sigma_{i}^{2}}}{\sum_{i}{ x_{i}^{2}/\sigma_{i}^{2}}}
\end{equation}
The uncertainty in $\mlb$ is

\begin{equation}
\label{eq:mlerr}
\sigma_{\mlb}= \frac{1}{\sqrt{(\sum_{i}{x_{i}^{2}/\sigma_{i}^{2}})}}
\end{equation}

The significance, estimated as the
strength of the zero-lag correlation relative to the rms found from the ensemble of randomized catalogs 
is $5.2\sigma$ (Table~\ref{tab:res}). 
That early type galaxy luminosity and total mass show such a strong correlation is the
central result of this paper.
The correlation strength at zero lag implies a $\mlb = 237 \pm 45 \hmlo$. The error is a 
$1\sigma$ statistical uncertainty only and does not include any systematic error 
introduced into $\scritinv$ due to lack of knowledge about the redshift distribution of the
source galaxies.

\subsection{Mass-Light Cross-Correlation Profiles }
\label{ssec:mlprof}

In order to determine how $\ml$ ratio varies with scale, we examine the profile of the
luminosity-mass cross-correlation and luminosity auto-correlation. 
Figure~\ref{fig:ccfp6} shows the azimuthally averaged profile of the 
mass-luminosity cross-correlation function from
(open circles with error bars) and the luminosity auto-correlation
function (filled circles). The luminosity auto-correlation has been normalized to 1, and in normalizing 
the mass-luminosity cross-correlation function, 
we have adopted a $\mlb = 250 \hmlo$, similar to that obtained at zero-lag in section~\ref{ssec:mlx}. The error bars were calculated from our ensembles of 32 noise reconstructions.
It is apparent from Figure~\ref{fig:ccfp6}
that although mass and luminosity do appear to trace each other, the profile is noisy, and thus it
is difficult to judge if the profiles have similar shape.

In the upper left panel of Figure~\ref{fig:ccfp4} we show the combined
azimuthally averaged profile  of the mass-luminosity cross-correlation 
function averaged over all six pointings
(open circles) and the corresponding luminosity auto-correlation function 
(filled circles). Again, the 
luminosity auto-correlation has been normalized to 1 and in normalizing 
the mass-luminosity cross-correlation function 
we have adopted a $\mlb = 250\hmlo$.

We see from Figure~\ref{fig:ccfp4} that the cross- and auto- correlation functions have very similar
profiles. Thus, it appears that early type galaxies trace the mass rather faithfully. We note that the profile
in Figure~\ref{fig:ccfp4} was obtained from smoothed images. The smallest scale over which 
mass/luminosity is being 
averaged is therefore $\simeq45\asec$. We conclude therefore that on 
scales $\geq45\asec$  our results are consistent with
early type galaxies tracing mass.

\begin{deluxetable}{cccrccr}
\tablewidth{0pt}
\tablecaption{$\mlb$ Values at Zero Lag.
\label{tab:res}
}
\tablehead{
\colhead{\omegamnought} &
\colhead{\omegalnought} &
\colhead{Lens Redshift} &
\colhead{Number Lens} &
\colhead{$\Sigma_{\rm crit}^{-1} (\times10^{-16} {\hMpc}^{2}\Mo^{-1})$}&
\colhead{$\mlb$} &
\colhead{$\nu$} 
\\
}
\startdata
\cutinhead{Flat Lambda}
\\
0.3 & 0.7 & $0.5\pm0.45$  & $5440$  & $1.91$  & $237\pm45$ & $5.2$
\\
\\
0.3 & 0.7 & $0.2\pm0.15$ & $680$   & $2.08$  & $353\pm67$ & $5.3$
\\
0.3 & 0.7 & $0.5\pm0.15$ & $3234$  & $2.10$  & $272\pm75$ & $3.6$
\\
0.3 & 0.7 & $0.8\pm0.15$ & $1526$  & $1.44$  & $-61\pm143$  & $-0.4$
\\
\\
\cutinhead{Einstein-de Sitter}
\\
1.0 & 0.0 & $0.5\pm0.45$  & $5018$  & $1.39$  & $294\pm60$ & $4.9$
\\
\\
1.0 & 0.0 & $0.2\pm0.15$ & $526$   & $1.72$  & $385\pm82$ & $4.7$
\\
1.0 & 0.0 & $0.5\pm0.15$ & $2966$  & $1.54$  & $376\pm101$ & $3.7$
\\
1.0 & 0.0 & $0.8\pm0.15$ & $1526$  & $0.98$  & $-93\pm211$  & $-0.4$
\\
\enddata
\end{deluxetable}

\subsection{Mass and Light as a Function of Redshift}
\label{ssec:interval}

In this section we investigate $\ml$, the constant of proportionality between mass and light, as 
a function of redshift.
We divide our data into three slices, each of size width $dz = 0.3$ (Table~\ref{tab:res})
and analyze each slice separately.

Figures~\ref{fig:pred_lock} to~\ref{fig:pred_16503} show the 
dimensionless surface mass density prediction $\kappa$-from-light generated as described in
Section~\ref{ssec:pred}. The remaining three panels in each figure show $\kappa$-from-light
summed over three redshift slices intervals centered on $z = 0.2$, $z = 0.5$, and $z = 0.8$
rather than the sum over all
nine redshift intervals described in \S~\ref{ssec:pred}.
We again assume a $\mlb = 300 \hmlo$ and use equations~\ref{eq:scritinv} and \ref{eq:kdomega}.
Luminosity is binned into equal angular-size pixels and we note that since our field of view corresponds to a 
different physical extent with redshift ($2.32\hMpc$ at $z = 0.1$; $9.82\hMpc$ at $z = 0.9$)
the binsize is similarly a function of redshift. It is apparent from Figures~\ref{fig:pred_lock} to~\ref{fig:pred_16503} that most of the $\kappa$-from-light signal 
originates at low and intermediate redshift. 


This effect can also been seen in Figure~\ref{fig:ccf3}. The left panels show the cross-correlation of
light with mass
averaged over all six pointings as a function of redshift. The right panels show the 
cross-correlation of light with an average over randomized shear catalog reconstructions. 
The slices
centered on redshifts $z = 0.2$ and $z = 0.5$ show a strong cross-correlation peak at zero lag.
The correlation strength at zero lag implies a $\mlb = 353\pm67\hmlo$ for  $z = 0.2$ with 
$5.3\sigma$ significance and $\mlb = 272\pm75\hmlo$ for  $z = 0.5$ with 
$3.6\sigma$ significance (Table~\ref{tab:res}).
The highest redshift slice shows no such peak, and a $\mlb$ consistent with zero ($-60\pm143\hmlo$).

Figure~\ref{fig:ccfp4} shows the azimuthally averaged profile of the 
mass-luminosity cross-correlation function from
(open circles with error bars) and the luminosity auto-correlation
function (filled circles). The redshift interval 
is marked on each panel. Again, 
the luminosity auto-correlation has been normalized to 1 but for better comparison of the profiles,
in normalizing 
the mass-luminosity cross-correlation function here
we have adopted a higher $\mlb = 300\hmlo$, more similar to that obtained at zero-lag
(rather than the $\mlb = 250\hmlo$ used for the whole sample in section~\ref{ssec:mlprof}). As before, uncertainties
were calculated from 32 reconstructions using randomly shuffled shear values from our $IV$ catalog.

There is perhaps some very slight evidence from Figure~\ref{fig:ccfp4}
that the mass-luminosity 
profile might be more extended than
the luminosity-luminosity profile. 
An extended mass-luminosity prtofile would be expected
if a) any mass associated with late type galaxies were non-negligible, and
b) one takes into account the known fact that
late type galaxies
are clustered around early types \markcite{dg-76}({Davis} \& {Geller} 1976) but are
more weakly clustered than early types around early types.
A more
realistic scenario, therefore, might include both early and late type
galaxies, with a slightly lower $\mlb$ for early types
than quoted here, and an additional much lower but non-zero $\mlb$ for late types
(the mass associated with late types cannot be dominant compared
to the early types or we would not see such a 
strong mass-to-early-type-luminosity
correlation in Figure~\ref{fig:ccf1}).
Further data will be required to determine conclusively whether the 
mass-luminosity profile is actually extended and not simply
a noise artifact. For now, however,  
we conclude, that within the 
uncertainties of this dataset,
we find no evidence for (early type galaxy) luminosity and (total)
mass segregation: it does appear that early type galaxies trace the mass very 
similarly at all redshifts.


We note from Figure~\ref{fig:ccfp4} that the uncertainties increase with increasing redshift. 
This is easily 
understandable when one realizes from equation~\ref{eq:mlrat} that, if the pointings have similar noise
properties as they do here, the mass-light correlation essentially consists 
of the product of 
$\kappa$-from-shear and $\kappa$-from-light  divided by the
product $\kappa$-from-light squared.

\begin{equation}
\label{eq:ml}
\mlb \rightarrow \frac{ \Sigma_{\rm pix} \kappa_{m} \kappa_{l}}{ \Sigma_{\rm pix} \kappa_{l} \kappa_{l}}
\end{equation}

Each factor of $\kappa$-from-light (equation~\ref{eq:kdomega})  contains a
factor of $\scritinv$ which at high redshifts becomes very small (Figure~\ref{fig:invscrit}). 
Thus for higher redshift lenses, one is dividing a fixed  
uncertainty in the numerator (the uncertainty in $\kappa_{m}$ due to intrinsic galaxy ellipticities)
by one factor of the decreasing quantity $\scritinv$,  and hence the uncertainty in the quotient, the mass-to-light ratio, increases
with redshift.
One could reduce the uncertainty in mass-to-light ratio at larger redshift if one had
deeper catalogs \ie\
catalogs containing source galaxies at higher redshift. Thus, one should not 
interpret Figures~\ref{fig:pred_lock} to \ref{fig:pred_16503}, and~\ref{fig:ccfp4} to \ref{fig:ccf3} as implying that there is negligible mass at redshift $z\geq0.8$, but rather, that without deeper catalogs one should not expect to be able to detect it. 

The value of $\mlb$ ($\simeq 300\hmlo$) obtained for the two lower redshift 
slices ($z = 0.2$ and $z = 0.5$) was higher than that obtained
for the combined sample. We interpret a depressed 
$\mlb$ value for the combined sample as a 
dilution effect due to the inclusion
of high redshift ($z = 0.6-0.9$) early 
types whose luminosity contributes to the auto-correlation (denominator)
 of equation~\ref{eq:ml} but whose mass does not
contribute to the numerator because of the reasons discussed in
the previous paragraph.

The $\mlb$ value of $\simeq 300\hmlo$ obtained from the two lower
redshift slices is therefore most likely to be representative
for this cosmology, and this is the value we shall adopt for the remainder
of the paper.


\section{DISCUSSION}
\label{sec:disc}

In the previous sections we used our CFHT data to cleanly select 
a sample of bright early type galaxies using $V-I$ colors and $I$ magnitudes, 
and assign reasonably precise redshifts to them.
We showed that there was a strong correlation between the actual mass inferred 
from weak lensing analysis and 
the mass predicted assuming the light associated with early type galaxies traces
mass with a constant mass-to-light ratio.

This is a surprising result which was first noticed for the ms0302 supercluster by KWLKGMD, as discussed
in the Introduction. The caveat,  in that case, of
course, was that as with all examples of \emph{cluster} measurements  
there was a strong possibility
of bias and no guarantee that a cluster $\ml$ ratio was representative of the Universe in general. There may have been
something unusual about the way that galaxies formed in such extremely overdense environments 
as clusters which was not
reproduced elsewhere under more ``normal'' conditions of formation. 
In this paper we have demonstrated the remarkable result that
early type galaxy light traces mass with constant $\ml$ ratio appears to 
be a universally applicable relationship. 

\subsection{$\mlb$ in an Einstein-de Sitter universe}
\label{ssec:cosmo}

The actual value of the constant of proportionality between mass and light is dependent upon cosmology. 
We performed our analysis assuming a 
flat lambda ($\Omega_{{\rm m}0} = 0.3, \Omega_{\lambda 0} = 0.7$) cosmology.
We also repeated the analysis for an Einstein-de Sitter 
($\omegamnought = 1.0, \omegalnought = 0.0$) cosmology.
The required $\mlb$ ratio \emph{increases} if this cosmology is assumed. 
Values of $\mlb$ at zero lag are shown in Table~\ref{tab:res}. The dependence of
$\mlb$ on cosmology can be understood qualitatively from the following argument.
From (\ref{eq:betaeff}) it is clear that the predicted $\kappa$-from-light is a function of
$\scritinv$. From Figure~\ref{fig:invscrit} it is clear that for an Einstein-de Sitter 
cosmology, $\scritinv$ is smaller than for the flat lambda case
at all redshift. Thus, the predicted $\kappa$-from-light in this cosmology is smaller by the same scaling
factor. Note that the scaling factor is not constant, but is a function of redshift, being the ratio 
of the height of the solid to dot-dashed line at each redshift. Hence, for this cosmology in order to 
match the same $\kappa$ fluctuations obtained from the $\kappa$-from-shear analysis, a 
\emph{higher} value of $\mlb$ ratio is required. 
 
\subsection{Comparison with other Galaxy Groups/Cluster Studies}
\label{ssec:clusml}

$\ml$ ratios have been measured on galaxy group and cluster scales by several other 
teams. On group scales, \markcite{hoek-99}Hoekstra, Franx, \& Kuijken (1999) recently found an average $\mlb =372\pm122$ for galaxy 
groups, after making a correction for luminosity evolution. On larger scales,
\markcite{mel-99}{Mellier} (1999, Table 1 and references therein) summarize all published $\ml$ ratios obtained for clusters using gravitational lensing as
of 1999. One should be cautious because the different teams used different telescopes, software
packages and mass reconstruction techniques, and obtained data under varying seeing and to varying physical radii from the
cluster center. Additionally, some teams choose to quote  $\mlv$ in preference to $\mlb$. Nevertheless, 
on scales of about 1 Mpc, the geometry of each cluster mass distribution inferred from gravitational lensing was similar to the galaxy
distribution, and also the X-ray distribution when it was available. The $\ml$ ratios 
obtained by different teams are fairly scattered, but have a median value of 300,
with a trend to increase with radius.  As clusters contain a high fraction of elliptical galaxies it is not
so unexpected that $\ml$ ratios for clusters turn out to be so similar to the value we derive
in this paper. 
 
\subsection{Inferred Mass Density \omegamnought}
\label{ssec:omegamnought}

We now calculate $\omegamnought$, the fractional contribution of early type galaxy mass density to the 
critical mass density  (where $\rho_{\rm crit}= 2.77\times10^{11} h^{2} \Mo {\rm Mpc}^{-3}$),

\begin{equation}
\label{eq:omegamnought}
\omegamnought =  \rho_{E}/ \rho_{\rm crit}
\end{equation}

\begin{equation}
\label{eq:rhom}
\rho_{E} =  (M/L)_{E}{\mathcal L}_{E} = (M/L)_{E} \phi_{\star E} L_{\star E} \Gamma(\alpha_{E}+2)
\end{equation}
where ${\mathcal L}_{E}$ is the measured $B$-band early type luminosity density of
the universe for \emph{early} type galaxies. With $\phi_{\star E}$ of $9\times10^{-3} (\hMpc)^{-3}$ , $L_{\star E}$ of $-19.61$, and $\alpha_{E}$ of $-0.74$ 
(from \markcite{folkes-99}{Folkes} {et~al.} 1999), we obtain ${\mathcal L}_{E}= 8.83\times10^{7}h\Lo {\rm Mpc}^{-3}$.

Using our $\mlb$ estimates for early types, and assuming a $\ml$ ratio of $300\hmlo$ ($400$ for Einstein-de Sitter) and an uncertainty of 
$\pm25\%$ (the uncertainty in our $\mlb$ value is a much greater contribution to the error budget than the 
uncertainty from 
\markcite{folkes-99}{Folkes} {et~al.}), we obtain $\omegamnought \simeq 0.10\pm0.02$  of closure density ($\omegamnought \simeq 0.13\pm0.03$ for Einstein-de Sitter).

The global density parameter we obtain appears low compared to other estimates \markcite{carl-97, mel-99}({Carlberg} {et~al.} 1997; {Mellier} 1999). 
This is not because our early type $\mlb$ value is low  
 but because we
do not assign the same $\mlb$ to late types as to early types. In the scenario we propose, late types are 
assumed to have very similar luminosities as early types (\eg\ \markcite{folkes-99}{Folkes} {et~al.} (1999) find $\Bstar$ for late types 
to be very similar to  $\Bstar$ for early types). The difference is that late types have much less 
mass associated with them and hence their $\mlb$ ratio is much lower. The analysis in this paper
assumes that they have negligible $\mlb$ compared to early types. 

Interestingly, we note that the current best limit on the baryon fraction determined from Big Bang
nucleosynthesis measurements, assuming $\Hnought = 65 \kmsMpc$, is $\Omega_{B} = 0.045\pm0.0028$ ($\Omega_{B}h^{2}= 0.019\pm0.0012$ from 
\markcite{burles-00}{Burles} {et~al.} 2000).
From (\ref{eq:betaeff}) and Figure~\ref{fig:invscrit} one would expect the $\mlb$ ratio for an 
 open baryon ($\omegamnought = 0.05$) universe $\simeq350\hmlo$ $\pm25\%$, intermediate between the
flat lambda and Einstein-de Sitter values. The lower limit to the mass fraction is thus
rather close to the baryon-only fraction suggesting that baryons might be the sole
source of mass in the universe. 
  A flat lambda cosmology with $\omegamnought = 0.05$ in baryons would
require an $\mlb$ ratio \emph{smaller} than that found for our fiducial $\omegamnought = 0.3$ case
which would in turn \emph{decrease} the estimates of  $\omegamnought \sim 0.10\pm0.02$ found above.
We also note in passing that preliminary analyses of recent
CMB measurements \markcite{jaffe-00}({Jaffe} 2000) prefer a higher value for $\Omega_{B} =0.076\pm0.012$
($\Omega_{B}h^{2} =0.032\pm0.005$)
than that derived by \markcite{burles-00}{Burles} {et~al.}. Of course, the upper limit we obtain for the
 mass fraction is triple the 
baryon-only fraction and would require sizeable quantities of exotic matter. Additionally, other
measurements, in particular supernovae constraints, suggest the Universe may contain
quantities of exotic matter.
Note also that all mass reconstructions from weak lensing analyses are 
blind to any uniform density component so in actuality, our $\mlb$ ratios 
should be considered to be lower limits.

\subsection{Possible Uncertainties}
\label{ssec:uncertain}

As a consistency check, we repeated the mass-luminosity correlation 
analysis in \S~\ref{sec:ml} but for $I$ and $V$
band catalogs separately. We obtained similar values  
of significance and $\mlb$ (within our 25\% uncertainties)
as a function of redshift as had been obtained for the  best composite catalog. 

The value of $\mlb$ we infer is strongly dependent on both cosmology and source
redshift distribution. Although we believe the effect of any uncertainty in the 
redshift distribution
of the source galaxies is largely dwarfed by intrinsic galaxy shape and measurement errors
we note that if the redshift distribution 
were in error and source galaxies were in fact at higher (lower)
redshift than our estimate, the inferred $\mlb $ ratio would decrease (increase).

Additionally, if there were some evolution in 
$\Lstar$ such that galaxies were brighter by a few tenths of a magnitude
between $z = 0.5$ and the present,
this would force an  upward revision to the $\mlb $ ratio inferred for the  $z = 0.5\pm0.15$ sample.
That the $z = 0.2\pm0.15$ and  $z = 0.5\pm0.15$ $\mlb$ values are so similar (Table~\ref{tab:res}) argues against strong luminosity evolution in the early types between redshift $0.5$ and the present. 

In the analysis performed in this paper we were forced by the availability of only
two passbands to select early type galaxies within $1-2$ magnitudes of $\Lstar$.  Presumably, 
$\mlb$ might not be constant with decreasing luminosity. However, the 
$\mlb$ ratio cannot increase dramatically at the  faint end of the luminosity function \ie\
faint galaxies cannot be very massive relative to their luminosity
or the correlation between mass and bright early type light seen in Figure~\ref{fig:ccf1}
would not be so convincing.
Also, there is of course, some mass associated with late type galaxies. Any mass associated with late type galaxies will contribute to the noise in Figures~\ref{fig:ccf1} and \ref{fig:ccf3}. 
However, as mentioned in \S\ref{ssec:interval} it cannot be dominant compared
to the early types or again, we would not see such a strong mass-to-early-type-luminosity
correlation.


Note that the values of $\mlb$ derived in this paper are entirely consistent with $\mlb$ ratios
for galaxy halos derived in WKLC from a galaxy-galaxy lensing analysis. In that paper, we found
a typical mass-to-light ratio of $\mlb \simeq 121\pm28h(r/100\hkpc)$ (for $\Lstar$ 
galaxies). In this paper, we found that halos extend to $\simeq 2 \amin$ at $z \sim 0.2$ (Figure~\ref{fig:ccfp4})
which corresponds to $r \sim 280 \hkpc$. 
Within that radius we therefore predict a  $\mlb$ of $2.8 \times 121\pm28\hmlo = 340\pm80\hmlo$
from that paper (assuming  a flat lambda ($\Omega_{{\rm m}0} = 0.3, \Omega_{\lambda 0} = 0.7$) cosmology) 
which compares well with the $\mlb$ of $ 300\pm75\hmlo$ derived from this paper.

We note, that in WKLC we adopted a different relationship between galaxy 
mass and light.  
On small scales ($\lesssim 10\hkpc$) it has 
been shown empirically that mass is approximately proportional to the square root of luminosity of early type galaxies \ie\ 
$M \propto \sqrt{L}$ \markcite{FJ-76, FT-91}(Faber \& Jackson 1976; Fukugita \& Turner 1991). In this paper we are probing scales larger
than individual galaxy halos, and it seems more reasonable to assume
$M \propto L$ on these scales. Incidentally, the 
$M \propto \sqrt{L}$
relationship on small scales  
also justifies our choice of smoothing
scale ($45\asec \sim 200 \hkpc$ at $z = 0.5$). A 
smoothing scale of a few hundred $\hkpc$ ensures we are investigating scales
``one step up'' from individual galaxy halos where 
$\ml$ biases on very small scales would tend to average out.

Based on the dataset described in this paper, it would be premature 
to claim definitively that (early type galaxy) luminosity traces mass with 
a \emph{constant} mass-to-light ratio.
The luminosity-luminosity and mass-luminosity profiles do appear remarkably 
similar. However, the uncertainties are sizeable
($\simeq 25\%$). We conclude that there is no 
evidence for (early type galaxy) luminosity and
mass segregation on galaxy group and cluster scales but that larger quantities of 
data will be required to 
determine empirically the mass-to-luminosity dependence to greater precision.  
That mass  should be directly proportional to luminosity
is a very appealing relationship but nature may have conspired otherwise,
and there may well still be scope
for some amount of biased galaxy formation.

\section{CONCLUSIONS}
\label{sec:conc}

Using $V-I$ color and $I$ magnitude we cleanly selected bright early type galaxies. We measured the
gravitational shear from faint galaxies and found a strong correlation with that predicted from the
early types if they trace the mass with $\mlb \simeq 300\pm75\hmlo$ for a flat ($\Omega_{{\rm m}0} = 0.3, \Omega_{\lambda 0} = 0.7$) lambda cosmology and
$\mlb \simeq 400\pm100\hmlo$ for Einstein-de Sitter.
We made two-dimensional reconstructions of 
the mass surface density. Cross-correlation of the measured 
mass surface density with that predicted from the
early type galaxy distribution showed a strong peak at zero lag. We azimuthally averaged the cross- and auto-correlation functions.
We concluded that the
profiles were consistent with early type galaxies tracing mass 
on scales of $\geq45\asec$ (the smoothing scale). 
We subdivided our bright early type galaxies by redshift and obtained similar conclusions.
These $\mlb$ ratios imply $\omegamnought \simeq 0.10\pm0.02$ ($\omegamnought \simeq 0.13\pm0.03$ for Einstein-de Sitter) of closure density.

In summary, we found that the majority of mass in the universe is associated with 
early 
type galaxies. On scales 
of $\geq 200\hkpc$ it appears that their light traces the underlying 
mass distribution with a constant $\mlb =300 -400 \pm 100 \hmlo$, depending on
cosmology.
As with several other recent results our data argues against an $\omegamnought=1$ universe.

In the future it will be possible to measure $\mlb$ ratios more precisely. The total areal coverage in
this paper was $1.5\deg^{2}$. Future planned surveys such as
the Hawaii Lensing Survey,
the Deep Lens Survey, or 
the Megacam/Terapix consortium 
will cover much larger area
and hence reduce uncertainties in $\mlb$. Additionally, more precise constraints on cosmology ($\omegamnought$
and $\omegalnought$) should soon be available from cosmic microwave background measurements, supernovae and the 
deep lensing surveys themselves. The color-redshift degeneracy could be broken by an increased number of passbands
to provide photometric redshifts. The greater range of absolute
luminosity then available (limited here to $L \sim L_* \pm 1-2$) would allow
mass-to-luminosity dependence 
(assumed in this work to be $M \propto L$) 
to be determined more precisely, both as a function of luminosity and as a function of distance from galaxy center. Finally, the availability of $>2$-passband data
would also allow photometric redshifts to be determined for late type galaxies and a similar 
investigation to be undertaken into their $\ml$ ratios.

\acknowledgements We thank Douglas Burke and Len Cowie for many 
useful discussions. We also thank Marc Davis, the referee, for
his constructive comments.
GW gratefully acknowledges financial support from the
estate of Beatrice Watson Parrent and from Mr. \& Mrs. Frank W. Hustace, Jr. whilst
Parrent Fellow at UH.  This work was supported by NSF grant
AST99-70805.




\clearpage

\begin{figure}
\caption[figs/kappa_lock_4_radec_forpaper.ps]{
Upper panels show reconstructions of mass surface density $\kappa$ $(= \scritrat)$ 
made from the $I$- and $V$-band 
catalogs separately for Lockman field (pointing 1). The lower left panel shows the
reconstruction from the composite $IV$ catalog, and the lower right panel shows the
reconstruction from the same catalog with randomized ellipticities, indicating the expected
noise fluctuations due to intrinsic random galaxy shapes. The reconstructions have been smoothed
with a $45\asec$ Gaussian filter.
The wedge shows the 
calibration of the grayscale and the contour separation is $0.04\times \scritrat$..
  
 \label{fig:k_lock}
}
\end{figure}

\begin{figure}
\caption[figs/kappa_lock_4_radec_forpaper.ps]{
Same as for Figure~\ref{fig:k_lock} but for Lockman field (pointing 2). 
 
\label{fig:k_lock2}
}
\end{figure}

\begin{figure}
\caption[figs/kappa_lock_4_radec_forpaper.ps]{
Same as for Figure~\ref{fig:k_lock} but for Groth field (pointing 1). 
 
\label{fig:k_groth}
}
\end{figure}

\begin{figure}
\caption[figs/kappa_groth_3_4_radec_forpaper.ps]{
Same as for Figure~\ref{fig:k_lock} but for Groth field (pointing 3). 
 
\label{fig:k_groth3}
}
\end{figure}

\begin{figure}
\caption[figs/kappa_1650_4_radec_forpaper.ps]{
Same as for Figure~\ref{fig:k_lock} but for 1650 field (pointing 1). 
 
\label{fig:k_1650}
}
\end{figure}

\begin{figure}
\caption[figs/kappa_1650_3_4_radec_forpaper.ps]{
Same as for Figure~\ref{fig:k_lock} but for 1650 field (pointing 3). 
 
\label{fig:k_16503}
}
\end{figure}

\clearpage

\begin{figure}
\caption[figs/ccf_FL_6_inv.ps]{
Cross-correlation of mass reconstruction from the $V$ catalog with mass reconstruction from the $I$ 
catalog (left panel) and with mass reconstruction from the randomized $I$ catalog (right panel)
for Lockman (upper) to 1650 (lower) fields.
 \label{fig:inv}
}
\end{figure}

\begin{figure}
\centering\epsfig{file=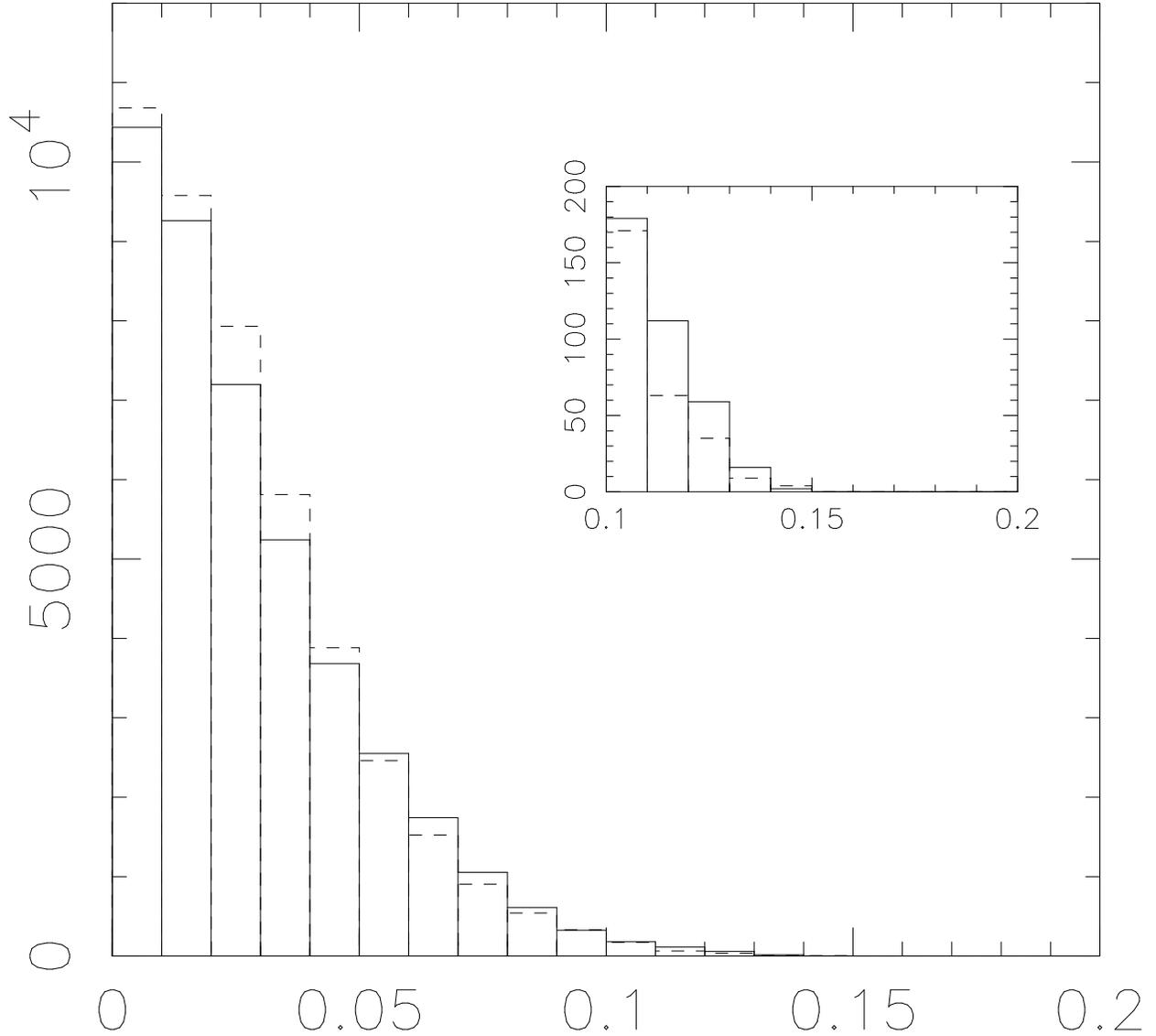,width=\figwidth}
\caption[hist.ps]{
Histograms of absolute pixel values from the reconstructions of mass surface density $\kappa$ 
using shear estimates from the $IV$ catalogs (Figures~\ref{fig:k_lock} to \ref{fig:k_16503}).
The solid histogram describes originally positive pixels and the dashed histogram describes
originally negative pixels. The inset shows the contrast for extreme values.
Clearly the distribution is very symmetrical. Notably, a highly positive tail is absent,
indicating the absence of very overdense structures \eg\ rich clusters.

 \label{fig:hist}
}
\end{figure}

\begin{figure}
\centering\epsfig{file=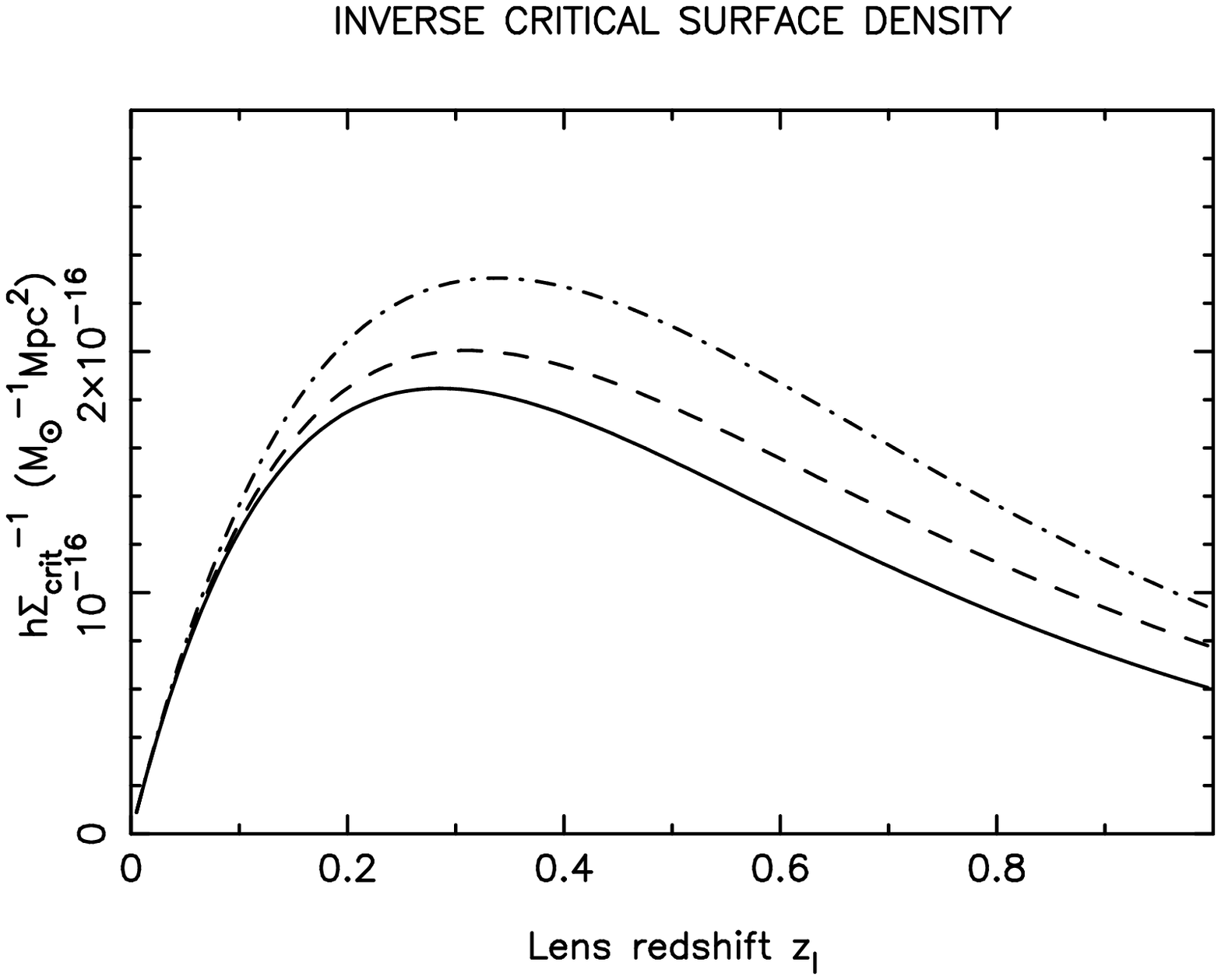,width=\figwidth}
\caption[plot_invsigmacrit_varys_i23.ps]{
Inverse critical surface density, $\scritinv$, as a function of redshift and
cosmology using the analytic approximation to an $m_{I} =23$ 
source galaxy redshift distribution. Solid line is 
Einstein-deSitter ($\omegamnought = 1.0$, $\omegalnought = 0.0$), dashed is
open baryon ($\omegamnought = 0.05$, $\omegalnought = 0.0$), dot-dashed is flat lambda ($\omegamnought = 0.3$, $\omegalnought = 0.7$).
  
 \label{fig:invscrit}
}
\end{figure}

\clearpage

\begin{figure}
\caption[figs/kappa_lock_4_radec_forpaper.ps]{
Upper left panel shows the predicted mass surface density using early type
galaxies selected by $V-I$ color, and $\mlb = 300\hmlo$ (see section~\ref{ssec:pred} for details) for Lockman field (pointing 1).  
The image has been smoothed with a $45\asec$ Gaussian filter.
The mean has been subtracted from the image. The wedge shows the 
calibration of the grayscale
and the contour separation is $0.007\times\scritrat$.
The remaining three panels show the predicted surface mass density using early type
galaxies and $\mlb = 300 \hmlo$ but subdividing the galaxies into 
$z = 0.2, 0.5, 0.8 \pm0.15$ . 

\label{fig:pred_lock}
}
\end{figure}

\begin{figure}
\caption[figs/kappa_lock_2_4_radec_forpaper.ps]{
Same as for Figure~\ref{fig:pred_lock} but for Lockman field (pointing 2). 
 
\label{fig:pred_lock2}
}
\end{figure}

\begin{figure}
\caption[figs/kappa_groth_4_radec_forpaper.ps]{
Same as for Figure~\ref{fig:pred_lock} but for Groth field (pointing 1). 
 
\label{fig:pred_groth}
}
\end{figure}

\begin{figure}
\caption[figs/kappa_groth_3_4_radec_forpaper.ps]{
Same as for Figure~\ref{fig:pred_lock} but for Groth field (pointing 3). 
 
\label{fig:pred_groth3}
}
\end{figure}

\begin{figure}
\caption[figs/kappa_1650_4_radec_forpaper.ps]{
Same as for Figure~\ref{fig:pred_lock} but for 1650 field (pointing 1). 
 
\label{fig:pred_1650}
}
\end{figure}

\begin{figure}
\caption[figs/kappa_1650_3_4_radec_forpaper.ps]{
Same as for Figure~\ref{fig:pred_lock} but for 1650 field (pointing 3). 
 
\label{fig:pred_16503}
}
\end{figure}

\clearpage

\begin{figure}
\caption[figs/ccf_FL_iv_1.ps]{
Cross-correlation of light with mass reconstruction (left panel) and 
with randomized catalog reconstruction (right panel). The contour separation is $1\times 10^{-6}$
and the peak at zero lag is significant at the $5.2\sigma$ level.
 \label{fig:ccf1}
}
\end{figure}

\begin{figure}
\centering\epsfig{file=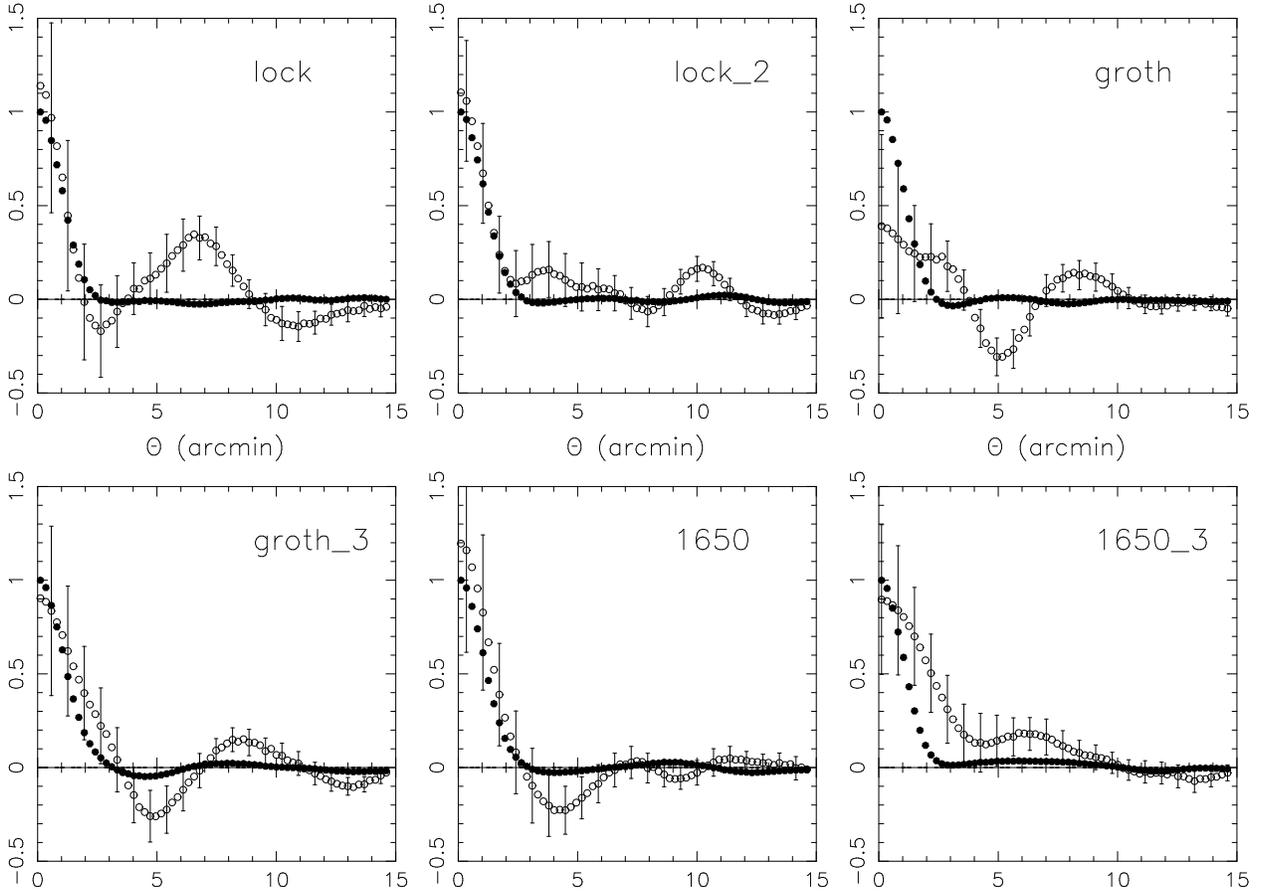,width=\figwidth}
\caption[ccfprofile_FL_6.ps]{
Azimuthally averaged profile of the mass-luminosity cross-correlation function 
(open circles with error bars) and the luminosity auto-correlation
function (filled circles). In normalizing the mass-luminosity cross-correlation function 
we have adopted a $\mlb = 250\hmlo$. For clarity, every third errorbar only is plotted. 
 \label{fig:ccfp6}
}
\end{figure}

\begin{figure}
\centering\epsfig{file=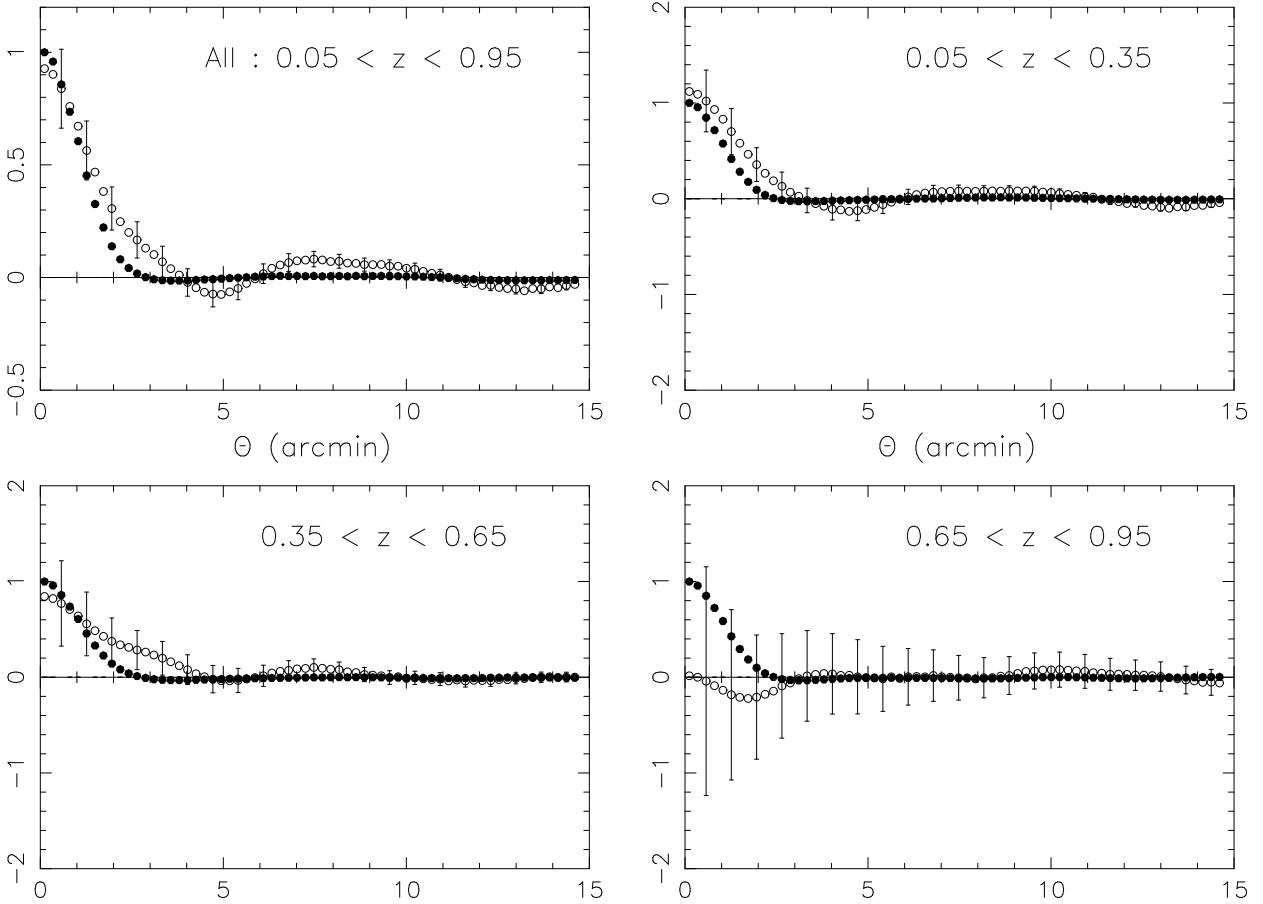,width=\figwidth}
\caption[ccfprofile_FL_1_4.ps]{
Azimuthally averaged profile of the mass-luminosity cross-correlation function from
Figure~\ref{fig:ccf1} (open circles with error bars) and the luminosity auto-correlation
function (filled circles). Upper left panel is for all galaxies (adopting a normalization $\mlb = 250\hmlo$), other 
panels are for
redshift intervals as marked (adopting a normalization $\mlb = 300\hmlo$). Note change of abscissa scale. For clarity, every third errorbar only is plotted. 
 \label{fig:ccfp4}
}
\end{figure}

\begin{figure}
\caption[figs/ccf_FL_iv_1_3n2.ps]{
As for Figure~\ref{fig:ccf1} but cross-correlation of light from galaxies in redshift 
intervals $z = 0.2, 0.5, 0.8 \pm 0.15$ 
with mass reconstruction (left panel) and 
with randomized catalog reconstruction (right panel). The contour separation is $5\times 10^{-7}$.
A correlation is seen between light and mass for galaxies at low and intermediate redshifts but 
no correlation is apparent
for galaxies in the highest redshift interval.
 \label{fig:ccf3}
}
\end{figure}

\end{document}